\documentclass[conference]{IEEEtran}

\usepackage{amsthm}
\usepackage{times,amsmath,epsfig}
\usepackage[T1]{fontenc}
\usepackage{graphicx} 
\usepackage{color} 
\usepackage{float}
\usepackage{boxedminipage}
\usepackage{multirow}

\usepackage{caption}
\usepackage{subcaption}

\usepackage{cite} 

\newtheorem{theorem}{Theorem}

\begin{document}

\title{Coded Slotted ALOHA with Varying Packet Loss Rate across Users}

\author{
\IEEEauthorblockN{
\v Cedomir Stefanovi\' c and Petar Popovski \\
}
\IEEEauthorblockA{
Aalborg University, Department of Electronic Systems,
Fredrik Bajers Vej 7, 9220 Aalborg, Denmark
\\
E-mail: \{cs,petarp\}@es.aau.dk
}
}

\maketitle

\begin{abstract}

The recent research has established an analogy between successive interference cancellation in slotted ALOHA framework and iterative belief-propagation erasure-decoding, which has opened the possibility to enhance random access protocols by utilizing theory and tools of erasure-correcting codes. 
In this paper we present a generalization of the and-or tree evaluation, adapted for the asymptotic analysis of the slotted ALOHA-based random-access protocols, for the case when the contending users experience different channel conditions, resulting in packet loss probability that varies across users. 
We apply the analysis to the example of frameless ALOHA, where users contend on a slot basis. We present results regarding the optimal access probabilities and contention period lengths, such that the throughput and probability of user resolution are maximized.

\end{abstract}

\section{Introduction}

Slotted ALOHA and its many variants are popular random access mechanisms used in various networks, e.g., cellular or satellite networks. 
The current rise of machine-to-machine (M2M) communications, with substantially increased number of terminals and different traffic patterns with respect to human-centric communications, requires efficient random access mechanisms and their redesign and enhancements are one of the main interests of M2M research community. 
An important research track in this direction is use of successive interference cancellation (SIC) in slotted ALOHA-based protocols \cite{CGH2007,L2011,SPV2012}. 
Particularly, the work presented in \cite{L2011} identified the analogies between SIC in framed slotted ALOHA and the iterative belief-propagation (BP) decoding of fixed-rate erasure-correcting codes, enabling the application of erasure-coding theory and tools.

The asymptotic analysis in terms of symbol erasure probability of erasure-correcting codes is standardly performed using and-or tree evaluation technique, first presented in \cite{LMS1998}, and applied in its original form for the slotted ALOHA case in \cite{L2011,SPV2012}.
However, both in \cite{L2011,SPV2012}, the underlying assumption is that the contending terminals are able to perform perfect power control, and thus experience the uniform probability of packet loss.
These results are not readily transferable to the case when link quality and packet loss vary across terminals. 

In this paper we derive the generalization of the and-or tree evaluation when the contending terminals experience different packet-loss probabilities. This assumption introduces fundamental differences with respect to the standard and-or tree evaluation used in erasure coding, where all input symbols contained (i.e., encoded) in a given output symbol have the same potential to contribute to the decoding process and there is no concept of varying quality across input symbols. In contrast, in a random access setting, the probability that a packet is successfully decoded depends on the channel conditions of the user sending the packet,
irrespective if there was a single user sending in the slot, or there were multiple users sending, but the packets of other users were removed with SIC.
The presented analysis can be applied to any slotted ALOHA-based method in which the user access strategy depends on the channel conditions.
We instantiate it for the case of \emph{frameless} ALOHA \cite{SPV2012}, providing insights on the relations between user access strategy, channel conditions, throughput and probability of successful resolution of user transmission, which hold in general for any coded slotted ALOHA access method.

\subsection{Related Work}

An extension of the and-or tree evaluation for the case of unequal error protection (UEP) rateless codes was assessed in \cite{RVF2007}, where input symbols are divided into classes with different selection probabilities when encoding the output symbol.
This was further generalized for the case of expanding-window LT codes \cite{SVDSP2009}.
Another extension was made for the case of distributed LT codes \cite{SPD2009}, where the encoding is performed independently by sources that can access only subsets (i.e., classes) of input symbols, and the outputs of the sources are combined by a relay.
This was developed further in \cite{TN2010}, where the relay, according to some probability distribution, choose either to forward or combine outputs of the sources that perform distributed encoding. 
Finally, a generalization of and-or tree evaluation where input and output symbols are divided into classes and selection probability of input symbols depend both on the output and input symbol class, was presented in \cite{SPDI2009}.
The presented work is closely related to \cite{SPDI2009}, as we also consider the case in which there are classes of users (i.e., input symbols), and slots (i.e., output symbols).
We assume that the division of the users into classes depends on the experienced channel conditions, which determines the user access strategy.
Our main contribution is the incorporation of the effects of unequal channel conditions into analysis; to the best of our knowledge, this has not been addressed in the previous work, as it is inherent to the random access framework.

\section{System model}
\label{sec:model}

\begin{figure}[tbp]
	\begin{center}
\includegraphics[width=0.55\columnwidth]{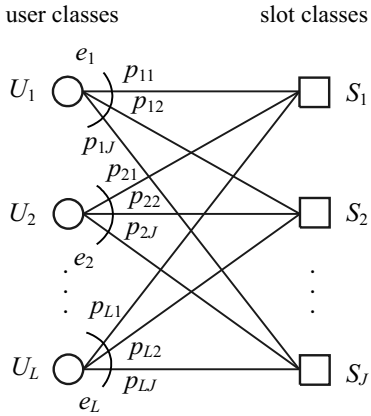}
	\end{center}
\caption{Graph showing connections among user and slot classes.}
	\label{fig:f-graph}
\end{figure}

We assume that there are $N$ contending users, divided into $L$ classes $U_l$, $l=1,...,L$, according to their probability of packet loss; the term \emph{packet loss} denotes the event when a user transmits interference-free and its packet is lost due to noise-induced errors.
Thus, the probability of packet loss depends only on the transmission power and channel conditions.
The packet loss probability for class $U_l$ is $e_{l}$ and a fraction of $a_l$ users belongs to $U_l$, $\sum_l a_l = 1$.
The length of the contention period is $M$ slots, where $M$, in general, does not have to be a priori fixed value.
The slots are divided into $J$ classes $S_j$, $j=1,...,J$, and fraction of $b_j$ slots belongs to class $S_j$, $\sum_j b_j = 1$.
Finally, the expected fraction of transmissions of class $U_l$ users taking place in class $S_j$ slots is given by $p_{lj}$; 
without loss of generality, we assume that:
\begin{align}
\label{eq:access_prob}
p_{lj} = \frac{\alpha_{lj}}{a_l N},
\end{align}
where $\alpha_{lj}$ are suitably chosen constants.
The above framework is depicted in Fig.~\ref{fig:f-graph}, showing connections among users and slots at class levels. We note that this framework is applicable both to framed ALOHA, where the users select transmission slot in a frame, and to slotted ALOHA, where the users decide whether to transmit or not on an individual slot basis.

Every user transmission consists of a replica of the same packet and a pointer to all other replicas.
The resolution of user transmissions is executed in the same way as iterative BP erasure-decoding.
First, slots containing a single packet (i.e., degree one slots) are identified and corresponding transmissions resolved.
In the next step, using pointers contained in the resolved packets, the slots containing the packet replicas are identified and these replicas are removed, potentially resulting with new degree one slots and instigating new iterations of the algorithm. 
We assume a perfect interference cancellation, i.e., once a user packet is resolved, all its replicas are perfectly removed.
This implies that whether a slot is useful or not when its degree is reduced to one, eventually depends on the packet-loss probability of the remaining packet.

\section{And-Or Tree Evaluation}
\label{sec:evaluation}

First, we briefly introduce the notation and terminology, following the standard introduced in \cite{LMS1998}.
For the general introduction on the and-or tree evaluation, we refer the interested reader to \cite{LMS1998,RU2007}. 
Denote a generic user from class $U_l$ by $u_l$ and a generic slot from $S_j$ by $s_j$.
By $\Lambda_{lj}^{(d)}$ denote the probability that user $u_j$ is connected to $d$ slots from class $S_j$, i.e., the probability that the degree of $u_j$ with respect to class $S_j$ is $d$.
The node-oriented degree distribution of $u_j$ with respect to $S_j$ is $\Lambda_{lj} (x) = \sum_{d=0}^{\infty} \Lambda_{lj}^{(d)} x^d$,
and the edge-oriented degree distribution of $u_l$ with respect to $S_j$ is
$\lambda_{lj} (x) = \frac{\Lambda'_{lj} (x)}{\Lambda'_{lj} (1)}$.
Similarly, by $\Omega^{(d)}_{jl}$ denote the probability that the slot $s_j$ is connected to $d$ users from class $U_l$, i.e., the probability that the degree of $s_j$ with respect to $U_l$ is $d$.
The node-oriented degree distribution of $s_j$ with respect to $U_l$ is $\Omega_{jl} ( x ) = \sum_{d=0}^{\infty} \Omega_{jl}^{(d)} x^d$, and the edge-oriented degree distribution of $s_j$ with respect to $U_l$ is $\omega_{jl} (x) = \frac{\Omega'_{jl} (x)}{\Omega'_{jl} (1)}$.
Denote the probability that the transmission of user $u_l$ is not resolved during the $i$-th iteration of the and-or tree evaluation by $y_l(i)$; by default we assume $y_l(0)=1$. 
The evaluation of $y_l(i)$ through iterations is given by the following theorem.
\begin{theorem}
\label{theorem}
\begin{align}
\label{eq:theorem}
& y_l(i)  = \nonumber \\
& \prod_{j} \lambda_{lj} \left( 1 - \sum_m \frac{\Omega'_{jm}(1)}{ \beta_{j}} ( 1 - e_{m}) \prod_k \omega_{jk} \left( 1 - y_k (i-1) \right) \right),
\end{align}
where $\beta_{j} = \sum_n \Omega'_{jn}(1)$ is the expected degree of $s_j$ and $i = 1,2,...$.
\end{theorem}

\begin{IEEEproof}
Assume that the degree of $u_l$ with respect to $S_j$ is $d_j$, for $1 \leq j \leq J$.
Using the standard ``or'' argument, it is easy to show that:
\begin{align}
y_l (i) = \prod_{j} r^{d_j}_{j}(i-1),
\end{align}
where $r_{j}(i-1)$ is the probability that a slot $s_j$ sends a not-resolved (i.e., erasure) message in the previous iteration.
Averaging over $d_j$, $1 \leq j \leq J$, yields:
\begin{align}
\label{eq:y}
y_l (i) = \prod_{j} \lambda_{lj} (r_{j} ( i - 1) ),
\end{align} 
where, with slight notation abuse, $y_l (i)$ also denotes the average.
Now, assume that the degree of $|s_j|$ with respect to $U_k$ is $d_k$, for $1 \leq k \leq L$.
The probability that a slot sends a not-resolved message, assuming that the last user that remains unresolved after SIC execution belongs to class $U_m$, is:
\begin{align}
\label{eq:not_resolved}
r_{j} (i-1) = 1 - ( 1 - e_{m}) \prod_k ( 1 - y_k (i-1) )^{d_k}.
\end{align}
where the product term stems from the standard ``and'' argument.
Averaging over $d_k$, $1 \leq k \leq L$, yields:
\begin{align}
r_{j} (i-1) = 1 - ( 1 - e_{m}) \prod_k \omega_{jk}( 1 - y_k (i-1) ),
\end{align}
where $r_{j} (i-1)$ is used to denote the average, as well.
Finally, averaging over all user classes with respect to the last unresolved user, \eqref{eq:not_resolved} becomes:
\begin{align}
\label{eq:r}
r_{j} (i-1) = 1 - \sum_m \frac{\Omega'_{jm}(1)}{ \beta_{j}} ( 1 - e_{m}) \prod_k \omega_{jk} ( 1 - y_k (i-1) ),
\end{align}
where $\frac{\Omega'_{jm}(1)}{ \beta_{j}}$ is the expected fraction of edges incident to $u_m$ that are also incident to $s_j$.
Substituting \eqref{eq:r} in \eqref{eq:y} completes the proof.
\end{IEEEproof}

The probability of resolving transmission of $u_l$ is simply $P_{Rl} = 1 - \lim_{i \rightarrow \infty} y_l (i)$.
The expected fraction of resolved users and the expected throughput, which are among key performance indicators of random access schemes, are:
\begin{align}
\label{eq:P_R}
P_R & = \sum_l a_l P_{Rl}, \\
\label{eq:throughput}
T  = & \frac{N P_R}{M}  = \frac{\sum_l a_l P_{Rl}}{ 1 + \epsilon},
\end{align}
where $\epsilon = \frac{M}{N} - 1$.

Going back to the system model, as depicted in Fig.~\ref{fig:f-graph}, it could be shown that $\Omega'_{jm}(1) = a_m N p_{mj} = \alpha_{mj}$, $\beta_j = \sum_n \alpha_{nj} $ and $\frac{\Omega'_{jm}(1)}{\beta_{j}} = \frac{\alpha_{mj}}{\beta_j}$.
Furthermore, we draw attention to another important fact, which generally holds for any slotted ALOHA-based scheme.
As users perform access independently and uncoordinatedly, slot degrees cannot be controlled directly, in contrast to the typical erasure-coding scenarios.
It is straightforward to show that the probability that the degree of $s_j$ with respect to $U_l$ is $d$ equals:
\begin{align}
\Omega^{(d)}_{jl} = { a_l N \choose d } p_{lj}^d ( 1 - p_{lj})^{ a_l N - d } \approx \frac{ \alpha_{lj}^d }{d!} e^{- \alpha_{lj}},
\end{align}
and the corresponding node-oriented and edge-oriented degree distributions are:
\begin{align}
\label{eq:O}
\Omega_{jl} ( x ) = \omega_{jl} (x) = e^{- \alpha_{lj} ( 1 - x )}.
\end{align}

Finally, we note that this generalization of the and-or tree evaluation is similar in flavor to the version of and-or tree lemma presented in \cite{LMS1998}, where the ``and'' nodes can be short-circuited to evaluate to zero with a predefined probability, irrespective to the values of input messages.
However, a crucial difference in the case assessed here is that the evaluation of ``and'' nodes to zero depends both on the packet-loss probability \emph{and} user access strategy through $\Omega'_{jm}(1)$, see \eqref{eq:theorem}, posing a different optimization problem.

\section{Results}

In this section we exemplify Theorem \ref{theorem} for the case of frameless ALOHA \cite{SPV2012}.
The key feature of frameless ALOHA is that users randomly and independently decide to transmit on a slot basis, using predefined slot access probability.
The slot access probability $p_{u_l \rightarrow s_j}$ depends both on the user class $U_l$ and slot class $S_j$, and is set to be equal to:
\begin{align}
p_{u_l \rightarrow s_j} = p_{lj} = \frac{\alpha_{lj}}{a_l N},
\end{align}
i.e., $p_{u_l \rightarrow s_j}$ is equal to the expected fraction of transmissions of class $U_l$ users taking place in class $S_j$ slots, see Section~\ref{sec:model} and Eq.~\eqref{eq:access_prob}.
The length of the contention period $M$ is not a priori set, and it lasts until a predefined criterion is satisfied; the termination criterion could be related to maximization/optimization of the fraction of resolved users \eqref{eq:P_R} and/or throughput \eqref{eq:throughput}.   
We note that the proposed access method was inspired by the encoding of rateless codes \cite{L2002,S2006}.
However, in contrast to the standard rateless coding, neither input \emph{nor} output distributions can be directly controlled in the random access framework. 

It is straightforward to show that in this case the probability that the degree of user $u_j$ with respect to class $S_j$ is $d$, is:
\begin{align}
\Lambda_{lj}^{(d)} & = { b_j M \choose d } p_{lj}^d ( 1 - p_{lj} )^{b_j M - d} \approx \frac{(b_j M p_{lj})^d}{d!} e^{- b_j M p_{lj}} \\
									 & = \left(  \frac{ b_j \alpha_{lj}}{a_l} \right) ^d\frac{(1 + \epsilon )^d}{d!} e^{- ( 1 + \epsilon ) \frac{b_j \alpha_{lj} }{a_l}}.
\end{align} 
Further, it could be shown that the node-oriented and edge-oriented degree distributions of $u_j$ with respect to $S_j$ are:
\begin{align}
\label{eq:L}
\Lambda_{lj} (x) = \lambda_{lj} (x) = e^{- ( 1 + \epsilon ) \frac{b_j \alpha_{lj} }{a_l} ( 1 - x )}.
\end{align}
The substitution of \eqref{eq:L} and \eqref{eq:O} into \eqref{eq:theorem} instantiates the theorem for the proposed random access method.

\begin{figure}[t]
        \centering
        \begin{subfigure}{0.95\columnwidth}
                \centering
                \includegraphics[width=\columnwidth]{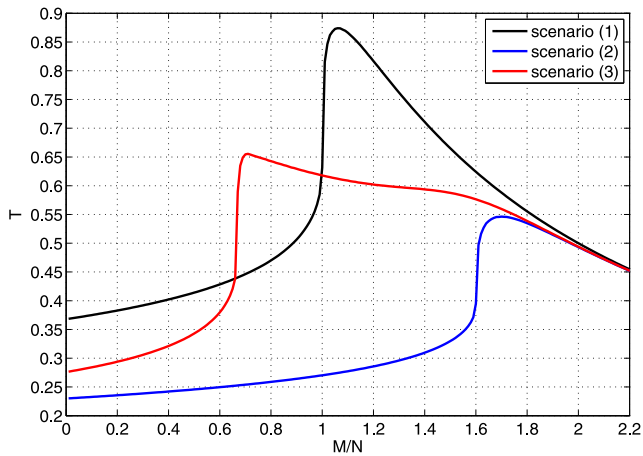}
                \caption{Throughput.}
                \label{fig:T}
        \end{subfigure}        
        \begin{subfigure}{0.95\columnwidth}
                \centering
                \includegraphics[width=\columnwidth]{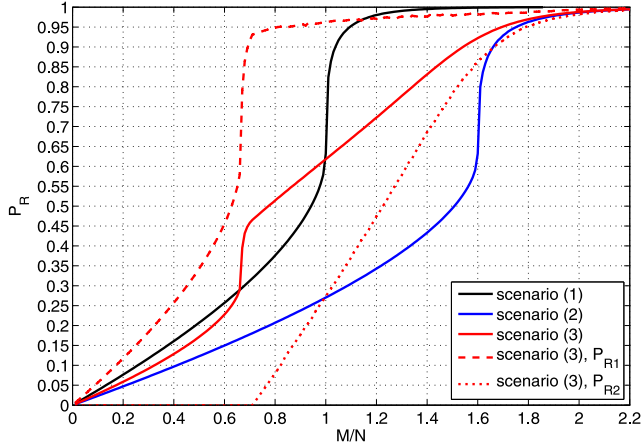}
                \caption{Probability of user resolution.}
                \label{fig:PR}
        \end{subfigure}  
        \begin{subfigure}{0.95\columnwidth}
                \centering
                \includegraphics[width=\columnwidth]{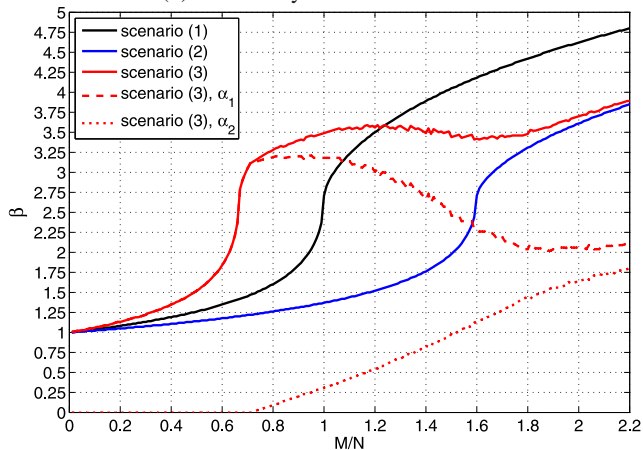}
                \caption{Expected slot degree.}
                \label{fig:beta}
        \end{subfigure}
   \caption{Results of the asymptotic for the given example.}\label{fig:example}  
\end{figure}

We apply the derived asymptotic analysis to the example consisting of the following scenarios:
(1) single user class and packet-loss probability $e^{(1)}=0$, (2) single user class and $e^{(2)}=0.375$, and (3) two user classes, equal fractions of user belonging to each class, i.e., $a_1^{(3)} = 0.5$ and $a_2^{(3)} = 0.5$, and  $e_1^{(3)} = 0.25$ and $e_2^{(3)}=0.5$.
In all cases we assume a single slot class; therefore, for scenarios (1) and (2) $\alpha_{11}^{(x)} = \beta_1^{(x)}$, $x=1,2$, and for scenario (3) $\alpha_{11}^{(3)} + \alpha_{21}^{(3)} = \beta_1^{(3)}$ (henceforth, we omit the subscripts corresponding to the user and slot classes when they are not required). 
Our aim is to find parameters $\alpha$ that asymptotically maximize the expected throughput.

Fig.~\ref{fig:example} depicts maximal expected throughput $T$ and corresponding maximal probability of resolution $P_R$ and optimal expected slot degree $\beta$, as functions of the ratio of the number of slots vs number of users $M/N$.
In other words, Fig.~\ref{fig:beta} shows what is the optimal $\beta$ that yields maximal $P_R$ and $T$ in Figs.~\ref{fig:PR} and \ref{fig:T}, for each value of $M/N$.

Comparison of the results for scenarios (1) and (2) reveals that, asymptotically, the overall maximal expected throughputs are $T_{max}^{(1)} \approx 0.87$ and $T_{max}^{(2)} \approx 0.55$, respectively; the corresponding lengths of the contention period are $M_{opt}^{(1)} \approx 1.05 N$ and $M_{opt}^{(2)} \approx 1.7 N$ slots, while $P_R^{(1)}=P_R^{(2)} \approx 0.93 $ and $\beta_{opt}^{(1)}=\beta_{opt}^{(2)} \approx 3.1 $.
These results are in line with a general reasoning that $\frac{T_{max}^{(2)}}{T_{max}^{(1)}} = \frac{M_{opt}^{(1)}}{M_{opt}^{(2)}} = \frac{1 - e^{(2)}}{1 - e^{(1)}}$, while $P_R$ and $\beta$ should be the same.

Further, the results corresponding to scenario (3) show that $T_{max}^{(3)} \approx 0.65$, which is obtained for $M_{opt}^{(3)}=0.7 N$, $\alpha_1^{(3)} \approx 3.1$ and $\alpha_2^{(3)} = 0$.
In other words, with respect to throughput maximization, the best strategy is to allow contention only among users with lower packet-loss probabilities (i.e., better channel conditions) and to silence the rest; this conclusion is also in line with expectations.
Following this insight, it could be shown that $\frac{T_{max}^{(3)}}{T_{max}^{(1)}} = \frac{1 - e_1^{(3)}}{1 - e^{(1)}}$ and that $\frac{M_{opt}^{(1)}}{M_{opt}^{(3)}} = \frac{1 - e^{(3)}}{a_1^{(3)}(1 - e^{(1)})}$.
On the other hand, if one aims to attain the overall high probability of user resolution $P_R^{(3)}$, which includes users belonging to the second class as well, this comes at the expense of lower expected throughput.
As $P_R^{(3)}$ increases and reaches $P_R^{(2)}$, $\alpha_2^{(3)}$ increases, but always remains lower than $\alpha_1^{(3)}$.
At the same time, $T_{max}^{(3)}$ tends to $T_{max}^{(2)}$, as now the average packet-loss probability of scenario (3) tends to packet-loss probability of scenario (2).

\section{Discussion and conclusions}

The presented analysis could be generally applied for the case when there is no power control, and its results used for a broader category of optimization problems, apart from the demonstrated throughput maximization.
For instance, one could further subdivide the users into classes with respect to the importance of their messages, and, following the presented methodology, derive and analyze the performance of the user access strategy.
Finally, we note that further generalizations of the and-or tree evaluation could include the impacts of imperfect SIC and capture effect, as outlined in \cite{L2011}.

\section*{Acknowledgement}

The research presented in this paper was partly supported by the Danish Council for Independent Research (Det Frie Forskningsr{\aa}d) within the Sapere Aude Research Leader program, Grant No. 11-105159 ``Dependable Wireless Bits for Machine-to-Machine (M2M) Communications'' and performed partly in the framework of the FP7 project ICT-317669 METIS, which is partly funded by the European Union. The authors would like to acknowledge the contributions of their colleagues in METIS, although the views expressed are those of the authors and do not necessarily represent the project.


\end{document}